# Optical Voltage Profiling of 2D Semiconductors via Proximal Exciton Sensing

Ha-Leem Kim[1,2,3,†,*], Hyungbin Lim[1,2,†], Yuanyi Yang[1], Ruishi Qi[1,2,3], Ruichen Xia[1], Can Uzundal[1,2,3], Takashi Taniguchi[4], Kenji Watanabe[5], Feng Wang[1,2,3,*]

[1]Department of Physics, University of California, Berkeley, CA 94720, USA.
[2]Materials Sciences Division, Lawrence Berkeley National Laboratory, Berkeley, CA 94720, USA.
[3]Kavli Energy NanoScience Institute, University of California, Berkeley and Lawrence Berkeley National Laboratory, Berkeley, CA 94720, USA.
[4]Research Center for Materials Nanoarchitectonics, National Institute for Materials Science, 1-1 Namiki, Tsukuba 305-0044, Japan.
[5]Research Center for Electronic and Optical Materials, National Institute for Materials Science, 1-1 Namiki, Tsukuba 305-0044, Japan.

*Correspondence to fengwang76@berkeley.edu; haleem4644@berkeley.edu

† These authors equally contributed to this work

## Abstract

High contact resistances in atomically thin semiconductors often mask intrinsic electrical transport properties, particularly at low carrier densities where exotic correlated states emerge. We introduce optical voltage profiling, a noninvasive wide-field technique that replaces local voltage probes with a proximal monolayer $MoSe_2$ exciton sensor. Isolated by thin hexagonal boron nitride, this sensor converts the target's local voltage into spatially resolved modulations of exciton reflectance. Through pixel-wise in situ calibration, these signals yield quantitative two-dimensional voltage maps of an actively biased semiconductor device. Using this method, we demonstrate the carrier-density-driven metal–insulator transition in bilayer $MoSe_2$ and obtain channel resistances of a few kΩ despite MΩ-scale two-terminal resistances in the metallic region. The optically derived resistance exhibits a metal–insulator crossover, and the voltage maps and reconstructed local conductivity reveal pronounced spatial heterogeneity in both insulating and metallic regimes. Beyond resolving channel resistance under high contact-resistance conditions, the technique provides spatially resolved access to microscopic transport heterogeneity in functional van der Waals devices.

## Introduction

Two-dimensional (2D) semiconductors provide a versatile platform for exploring quantum many-body states, aided by reduced dimensionality and gate-tunability. Transition-metal dichalcogenides (TMDs) are particularly attractive owing to their high-quality crystals, tunable bandgaps, and strong excitonic effects[1,2]. At low carrier densities, electron–electron interactions dominate over kinetic energy, giving rise to Wigner crystals[3,4] and fractional quantum Hall states[5,6]. Heterostructures and moiré superlattices further extend this landscape through band engineering, enabling access to even more exotic phases such as superconductors[7,8], Chern insulators[9], and fractionally quantized anomalous Hall states[10,11].

Progress, however, is often limited by electrical contacts. Schottky barriers at metal–TMD interfaces routinely lead to megaohm (MΩ) resistances that overwhelm channel signals[12], especially at low electron densities where quantum effects are most pronounced. Conventional four-probe measurements can partially mitigate the contact resistance, but an ohmic contact is usually needed. However, achieving such ohmic contacts reproducibly across materials and devices remains challenging.

A variety of approaches have been explored to reduce contact resistance—including interfacial charge-transfer doping[13], van der Waals semimetal electrodes[14,15], transferred metals[16], edge contacts[17] and surface treatments[18]. Despite important case-by-case successes, no universal recipe has been developed: performance depends sensitively on the metal–semiconductor pair, interface chemistry and alignment, and cannot be predicted by work-function matching alone. This lack of generality continues to restrict systematic studies of emerging 2D materials at low carrier densities.

Here we introduce a noninvasive optical voltage-profiling technique that circumvents the contact bottleneck. A monolayer TMD sensor (1L-$MoSe_2$) is placed adjacent to the target layer, with the two layers separated by thin hBN. When current flows through the target, its local electrostatic potential modulates the carrier density in the sensor and shifts the excitonic resonance. With a calibrated response, changes in the sensor's exciton intensity translate directly into the local voltage of the target. Using wide-field illumination, we obtain full two-dimensional voltage maps rather than discrete readouts at a few points. We validate this approach on bilayer $MoSe_2$ and investigate the metal–insulator transition (MIT). Despite two-terminal resistances on the order of MΩ, we resolve channel resistances of a few kΩ in 2L-$MoSe_2$. We further extract the channel resistance as a function of carrier density and temperature and observe a metal–insulator crossover. The crossover density obtained optically agrees with that measured on a separate device by conventional four-probe transport, providing an independent consistency check of the extracted critical density.

## Results

### Voltage profiling principles and device architecture

The optical voltage-profiling technique mirrors a four-probe measurement but replaces voltage leads with light, requiring only source and drain contacts to the target layer (Fig. 1a–c). The heterostructure consists of a target layer (the material under study; here 2L-$MoSe_2$) and a sensor layer with a sharp, carrier-sensitive optical resonance (here 1L-$MoSe_2$). Monolayer TMDs such as $WSe_2$, $MoSe_2$, and $WS_2$ are well-suited sensors owing to their narrow exciton linewidths. The layers are separated by thin hBN (3–4 nm) to enable strong electrostatic coupling.

The sensing mechanism relies on the target acting as a local gate for the sensor[19,20]. A source–drain voltage applied to the target induces a position-dependent voltage, which shifts the interlayer bias and thereby the local Fermi level in the sensor. This changes the balance between the neutral exciton $X_0$ and the trion $X_-$, providing an optical transduction of the target's voltage. In the unperturbed state (zero source–drain bias), the interlayer bias $V_{bias}$ is set at the point of maximum responsivity—near the sensor's charge-neutral to electron-doped transition (Fig. 1a). Reflection contrast spectra versus $V_{bias}$ resolve the A-exciton ($X_0$) and trion ($X_-$) peaks in 1L-$MoSe_2$ (Fig. 1d), with the optimal bias marked by a white dashed line.

Under finite source–drain voltage, current flow generates a spatial voltage profile with pronounced drops at the contacts—reflecting large contact resistance—and a gradient across the channel (Fig. 1b). Near the source, electron doping suppresses $X_0$ and enhances $X_-$; near the drain, the opposite occurs. Position-dependent reflectance changes thus encode the target's local voltage. Focused beams provide discrete readouts[20] (Fig. 1b), whereas wide-field illumination yields full two-dimensional maps (Fig. 1c). For wide-field imaging we employ incoherent Bessel illumination generated by an axicon and a rotating diffuser to suppress speckle[21], and we confine the illumination to the region of interest to limit optical power to < 60 nW. Illumination at 754 nm targets the 1L-$MoSe_2$ $X_0$ resonance to enhance responsivity.

To probe the density-dependent resistance of 2L-$MoSe_2$, the device geometry incorporates graphite ribbon contacts for reduced contact resistance via increased contact length (see Supplementary Note 1). Three gates provide controlled doping. The back gate and contact gate are biased positively, approaching 0.5 V $nm^{-1}$ to heavily electron-dope the contact regions in both 1L and 2L layers (Fig. 1e). The fixed interlayer bias sets the 1L-$MoSe_2$ at the maximum responsivity, while the top gate selectively depletes 2L-$MoSe_2$ electron density in the channel—leaving contacts metallic—to drive the metal–insulator transition. This configuration mitigates current crowding, as the heavily doped regions spread current uniformly from contacts to the channel. Source–drain voltage is applied using a dual-channel function generator, while current is measured with a lock-in amplifier via voltage drop across a series resistor (10 kΩ) connected to the source terminal. An optical microscope image of a representative device is shown in Fig. 1f.

## Gate-dependent doping in the sensor and target layer

To probe transport phenomena in the target heterostructure using optical voltage profiling, we first mapped the gate-dependent doping behavior of both 1L-$MoSe_2$ (sensor) and 2L-$MoSe_2$ (target). Data in Fig. 2 are acquired from device DV1, with an interlayer hBN thickness of 3.3 nm and a top-gate hBN thickness of 8 nm; all measurements were performed at 6.5 K. With the back and contact gates fixed at high positive voltages to keep the contacts metallic, the remaining control variables are the top gate voltage $V_{TG}$ and the interlayer bias $V_{bias}$. The doping behavior of the 1L-$MoSe_2$ can be determined from its exciton ($X_0$) intensity as a function of $V_{TG}$ and $V_{bias}$ (Fig. 2a), where strong $X_0$ intensity indicates the charge-neutral regime, and reduced $X_0$ intensity corresponds to electron doping. The 2L-$MoSe_2$ electron density change is reflected through its two-terminal conductance (Fig. 2b): the conductance becomes nonzero when the 2L-$MoSe_2$ becomes doped and continues to increase with increased electron density. Note that the 1L $X_0$ map also shows a diagonal boundary that coincides with the charge-neutrality line of the target (black dashed line in Fig. 2b). This boundary can be explained by electrostatic screening of the top gate by the target, which lies between the top gate and the sensor (Fig. 1e). When the target is electron-doped, its free carriers screen the top-gate field, and the sensor density depends on $V_{bias}$ alone. When the target is charge-neutral, the top-gate field penetrates the target and also depletes the sensor, so that the 1L $X_0$ intensity increases as $V_{TG}$ becomes more negative.

These maps allow us to choose a trajectory that maintains the sensor at maximum responsivity while sweeping the target electron density across the metal–insulator transition. We fix $V_{bias} = 0.81$ V—near the 1L-$MoSe_2$ doping threshold for peak sensitivity—and vary $V_{TG}$ to control the 2L-$MoSe_2$ electron density (white dotted line in Fig. 2b). A line cut of the two-terminal conductance along this trajectory is shown in Fig. 2c, which first increases with the carrier density and then saturates at around 1 μS. The saturation of the conductance, which corresponds to a two-terminal resistance of ~1 MΩ, is likely due to the contact resistance.

## Voltage reconstruction from in situ calibration

Optical voltage readout is conceptually straightforward but experimentally subtle. Local interlayer-bias changes determine the sensor carrier density via the interlayer capacitance; however, the conversion from local carrier density to the exciton resonance intensity is nonlinear and can vary with spatial location due to sample inhomogeneity.

To account for these variations, we implement an in situ, spatially resolved calibration (Fig. 3a,b). Each measurement period consists of two phases. In the calibration phase, the source and drain are set to the same calibration voltage $V_{\mathrm{cal}}$. Here and below, the voltage $V$ of the target denotes the electrochemical potential of its electrons divided by $-e$, $V = \varphi - \mu/e$, where $\varphi$ is the electrostatic potential and $\mu$ is the chemical potential. $V$ is the quantity read by a voltage probe in a four-probe measurement. With no source–drain current, the target equilibrates to a uniform $V = V_{\mathrm{cal}}$, even where its conductivity, $\varphi$ and $\mu$ vary with position. In the biasing phase, a small differential voltage is applied (typically +30 mV at the source and −30 mV at the drain). The camera is triggered once per phase to acquire a pair of images per period; frames for each phase are averaged over multiple periods to improve signal-to-noise ratio. For balanced contacts, where the voltage drops at the source and drain are similar, the channel voltage during the biasing phase remains near 0 V relative to ground (Fig. 3b). For imbalanced contacts, where the voltage drops at the source and drain contacts are very different, a DC offset will be added to both contacts during the biasing phase so that the channel voltage remains centered near 0 V while the source–drain voltage difference is held fixed. The required offset varies with the channel carrier density, since the degree of contact imbalance itself depends on the channel resistance.

We then form the digital lock-in signal by subtraction, $\Delta I(V_{\mathrm{cal}}, x, y) = I_{\mathrm{cal}}(V_{\mathrm{cal}}, x, y) - I_{\mathrm{bias}}(x, y)$, where $I_{\mathrm{cal}}(V_{\mathrm{cal}}, x, y)$ is the reflectance image acquired during the calibration phase at voltage $V_{\mathrm{cal}}$ and $I_{\mathrm{bias}}(x, y)$ is the reflectance image acquired during the biasing phase. To extract the local voltage, $V_{\mathrm{cal}}$ is swept over a narrow window (±20 mV). With the gates and the sensor held at fixed voltages, the local carrier densities of the target and the sensor, and hence the local reflectance, are determined by the local target voltage. For a given position, $\Delta I(V_{\mathrm{cal}})$ crosses zero when the calibration voltage equals the local target voltage during the biasing phase. Thus, the zero-crossing of $\Delta I(V_{\mathrm{cal}})$ yields the local voltage of the target layer. Fig. 3c shows the digital lock-in signal versus the swept $V_{\mathrm{cal}}$ at a representative pixel. We determine the zero-crossing by a linear fit of $\Delta I(V_{\mathrm{cal}})$ within ±3 mV of the crossing. The noise of a single calibration step (shaded band) corresponds to 0.4 mV, and the fit over the calibration steps in this window gives a voltage resolution of about 0.1 mV per pixel (Supplementary Note 2). Each camera pixel corresponds to 0.37 μm on the sample, finer than the spatial resolution of about 1 μm set by the diffraction limit of the objective (NA = 0.4) at λ = 754 nm. Applying the zero-crossing analysis pixel-wise reconstructs the full two-dimensional voltage map. Figure 3d displays a representative voltage map of a 2L-$MoSe_2$ device (DV1) at a bias current of 100 nA and an electron density of $1.0 \times 10^{12}$ cm$^{-2}$. The voltage profiling works well in the area where the target (2L-$MoSe_2$) and sensor (1L-$MoSe_2$) layers overlap. Notably, DV1 exhibits substantial electrical transport inhomogeneity, evidenced by the complex equipotential contours that are not perpendicular to the source–drain axis. (This very strong inhomogeneity makes it difficult to extract the channel resistance reliably.)

The optical voltage profiling technique with in-situ calibration is operationally robust. Because calibration and biasing images are acquired within the same period, pairwise subtraction cancels common-mode fluctuations (illumination and mechanical drift), and residual short-term fluctuations are suppressed by averaging images over multiple periods. Moreover, acquiring a full spatial map each cycle makes the extraction of voltage differences between two positions insensitive to long-term sample or beam drift.

## Density-dependent metal–insulator transition

For a quantitative analysis of the density-dependent channel resistance, we focus on a narrower device DV2, whose voltage maps show a more regular voltage drop along the channel than those of DV1, although spatial inhomogeneity persists. Fig. 4a shows the spatial maps of the local voltage $V(x, y)$ for device DV2 at a source–drain bias voltage of 60 mV and T = 6.5 K when the carrier density is varied from $0.40 \times 10^{12}$ to $1.21 \times 10^{12}$ $cm^{-2}$. The corresponding bias current is shown in the inset of Fig. 4h. The overall voltage drop across the device decreases as the electron density increases, reflecting the reduction of channel resistance. As seen in the measured voltage maps (Fig. 4b, e), the equipotential contours remain notably irregular even at the highest density, indicating that spatial inhomogeneity in the channel persists into the metallic regime. To further examine this behavior, we reconstruct the local conductivity and current distribution at one low electron density point (n = $0.40 \times 10^{12}$ $cm^{-2}$, insulating) and one high electron density point (n = $1.21 \times 10^{12}$ $cm^{-2}$, metallic).

We reconstruct the local conductivity $\sigma(x, y)$ by solving an inverse problem constrained by charge conservation and Ohm's law, $\nabla \cdot [\sigma(x, y)\nabla V(x, y)] = 0$, using the measured voltage maps (Fig. 4b, e) as input. Mild Gaussian smoothing is applied to these maps prior to inversion to suppress high-frequency noise that would otherwise destabilize the algorithm. The inversion is performed within a finite-element framework. Since multiple $\sigma(x, y)$ distributions can reproduce the same voltage map, we apply regularization that penalizes abrupt, unphysical variations in $\sigma$ to converge to a physically plausible solution (Supplementary Note 3). The resulting conductivity maps are shown in Fig. 4d, g. At low density (Fig. 4d), the overall conductivity is small and highly inhomogeneous across the device. At high density (Fig. 4g), the conductivity is significantly larger, but its spatial map remains visibly inhomogeneous.

Using the reconstructed $\sigma(x, y)$, we recompute the voltage and evaluate the local current density $\mathbf{J} = -\sigma\nabla V$ (Fig. 4c, f). The recomputed voltage reproduces the measured voltage map (compare Fig. 4b, e with Fig. 4c, f), confirming the internal consistency of the inversion algorithm. At both densities, the current preferentially flows through the high-$\sigma$ regions and avoids the low-$\sigma$ ones, so transport is never strictly uniform. Pronounced lateral redirection of current is clearly visible at both low and high densities, demonstrating that channel inhomogeneity persists across the entire density range, including deep in the metallic regime.

We extract the average channel resistance $R = \Delta V/I_{SD}$ in analogy with a conventional four-probe measurement, but with $\Delta V$ obtained optically: $V(x, y)$ is averaged over two interior regions near the source and drain contacts (indicated on the $n = 1.21 \times 10^{12}$ $cm^{-2}$ map in Fig. 4a), and $\Delta V$ is the difference of the two regional averages. The resulting sheet resistance $R_{\square}(n)$ at $T = 6.5$ K is plotted in Fig. 4h, with the corresponding $\Delta V$ and $I_{SD}$ shown in the inset. With the top gate and the sensor held at fixed voltages, a higher (lower) local target voltage lowers (raises) the local electron density of the target. The gray band in Fig. 4h shows the bias-induced density range between the two averaging regions (Methods). It becomes comparable to $n$ below $0.3 \times 10^{12}$ $cm^{-2}$, where $R_{\square}$ represents an average over a broad range of densities.

At low density, the density gradient induced by the source–drain bias can also contribute to the inhomogeneity of the reconstructed conductivity (Fig. 4d). At $n = 0.40 \times 10^{12}$ $cm^{-2}$, the voltage varies by about ±10 mV along the channel, which corresponds to a density variation of about $\pm 0.08 \times 10^{12}$ $cm^{-2}$ (gray band in Fig. 4h). Given the density dependence of $R_{\square}$ in Fig. 4h, this variation lowers the conductivity on the high-voltage side and accounts for a conductivity change of up to about a factor of three along the channel. The reconstructed conductivity in Fig. 4d, however, varies by about an order of magnitude along the channel and also varies across it. At $n = 1.21 \times 10^{12}$ $cm^{-2}$, the density variation is below $0.01 \times 10^{12}$ $cm^{-2}$, yet the conductivity remains visibly inhomogeneous (Fig. 4g). The bias-induced density gradient could therefore contribute to the inhomogeneity at low density, but it alone does not account for the full extent of the observed inhomogeneity.

To identify the critical density from the channel-averaged $R_{\square}$, we plot temperature-dependent $R_{\square}(T)$ traces from $n = 0.5$ to $1.2 \times 10^{12}$ cm$^{-2}$ in Fig. 4i. At $n \sim 0.5 \times 10^{12}$ cm$^{-2}$, the resistance first decreases on cooling, passes through a minimum (the "valley"), and then rises again on further cooling. This valley behavior can be understood from the Drude expression $\rho \propto 1/(n\tau)$ by considering the temperature dependences of $n$ and $\tau$ separately. At low temperature, carriers freeze out following $n \propto e^{-\Delta/k_BT}$, so $\rho \propto e^{\Delta/k_BT}$ — the activated insulating regime. At high temperature, all carriers are thermally activated and phonon scattering gives $\tau \propto 1/T$, leading to $\rho \propto T$ (metallic-like). The valley arises from the crossover between these two regimes. The critical density of the metal–insulator transition is usually defined by a temperature-independent resistivity as $T \rightarrow 0$. Here we take $n_{\mathrm{MIT}} \sim 7 \times 10^{11}$ cm$^{-2}$ as the density at which $\mathrm{d}R/\mathrm{d}T \sim 0$ for the two lowest-temperature traces. The bias-induced density spread near $n_{\mathrm{MIT}}$ is about $\pm 0.03 \times 10^{12}$ cm$^{-2}$, small compared with $n_{\mathrm{MIT}}$ itself.

To validate the optically extracted resistance, we fabricated multiple 2L-$MoSe_2$ devices for conventional four-probe transport measurements. Only a very small fraction of the 2L-$MoSe_2$ transport devices is functional with sufficiently low contact resistance. The transport data for one functional device are shown in Fig. 4j, which agrees qualitatively with the results obtained through optical voltage profiling. Specifically, both types of measurement yield a critical density $n_{\mathrm{MIT}} \sim 7 \times 10^{11}$ cm$^{-2}$, but they differ in the absolute value of R. At $n_{\mathrm{MIT}}$, $R_{\square}$ is $\sim 30$ kΩ from optical voltage profiling measurement, while the transport value is $\sim 6$ kΩ. The higher resistivity at a given doping in the optical voltage profiling data might originate from the sample inhomogeneity. The heterostructure used in optical measurements contains more than ten transferred layers, compared with three in the transport device, and accumulates more interfacial bubbles and strain during fabrication (dashed circles in Fig. 4a). If the current flows mainly through a narrow conducting path, $n_{\mathrm{MIT}}$ is set by this path, whereas $R_{\square}$ is calculated with the full channel width. A phenomenological simulation shows that such a path leaves $n_{\mathrm{MIT}}$ unchanged and raises the critical $R_{\square}$ by about the ratio of the channel width to the path width (Supplementary Note 4). This interpretation is consistent with the nonuniform current flow and the strong inhomogeneity of the reconstructed conductivity in Fig. 4c, d, f, g.

## Discussion

In conclusion, we introduce optical voltage profiling as a noninvasive wide-field technique that uses a proximal monolayer exciton sensor to map internal voltages in 2D devices using only source–drain contacts. Through pixel-wise in situ calibration, the method yields quantitative voltage maps robust to exciton spatial inhomogeneity, sample drift, and contact asymmetry.

Applied to bilayer $MoSe_2$, it resolves channel resistances of a few kΩ despite MΩ-scale two-terminal resistances, revealing a density-driven evolution from insulating to metallic conduction. Reconstruction of the local conductivity and current flow from the voltage maps shows that conductivity inhomogeneity persists from the insulating regime well into the metallic regime, with current preferentially routed through higher-conductivity domains throughout. The optically extracted average resistance exhibits a metal–insulator crossover at a critical density that agrees with conventional four-probe transport.

Beyond solving the contact bottleneck, optical voltage profiling expands transport metrology in quantum 2D systems. By imaging internal voltages along the current path, it avoids geometric mixing and enables quantification of intrinsic anisotropy. Access to internal voltages without perturbing edges facilitates separation of bulk and edge conductance in edge-dominated topological materials such as monolayer $WTe_2$[22-24] and $MnBi_2Te_4$[25]. The technique is compatible with various excitonic sensors, operates at low optical power and cryogenic temperatures, and integrates naturally with dual-gate device structures. It provides a powerful tool for exploration of low-density and topological phases in van der Waals heterostructures.

## Methods

### Device fabrication

Monolayer and bilayer $MoSe_2$, few-layer graphene, and hexagonal boron nitride (hBN) were mechanically exfoliated onto $SiO_2/Si$ substrates. A 3–5 nm hBN spacer was used between the 2L-$MoSe_2$ target and the 1L-$MoSe_2$ sensor. Graphite contacts to the 2L-$MoSe_2$ were patterned into a comb geometry by atomic-force-microscope (AFM) lithography[26] to increase the effective edge length at the $MoSe_2$/graphite interface and thereby lower contact resistance. Heterostructures were assembled using polycarbonate (PC) film stamps, and the completed stack was released onto a $SiO_2/Si$ substrate with pre-patterned markers at 160 °C. The contact-gate electrode was defined by electron-beam lithography followed by metal evaporation (Cr/Au, 5 nm/60 nm). Bond pads for wire-bonding were patterned by photolithography and metal evaporation (Cr/Au, 5 nm/120 nm).

### Optical measurement

All optical measurements were performed in a variable-temperature cryostat (Montana Instruments). Top-gate, back-gate, contact-gate, and interlayer-bias voltages were applied using source meters (Keithley 2400/2612).

**Reflection-contrast spectroscopy:** A supercontinuum laser (YSL Photonics) or a diode laser operated in LED mode served as the light source. Spectra were collected with a nitrogen-cooled CCD camera. The excitation power at the sample was kept $\leq$ 1 nW.

**Optical voltage profiling measurement**: A diode laser operated in lasing mode was temperature-tuned to the 1L-$MoSe_2$ A-exciton resonance ($\approx$ 754 nm) to maximize responsivity. The exciton of the 2L-$MoSe_2$ target lies at about 766 nm, more than 10 nm from the probe wavelength (Supplementary Note 5). Output power was stabilized with a Noise Eater (Thorlabs NEL03A). For two-point (discrete) voltage readout, non-polarizing beam splitters divided the beam into two paths that were focused onto two independent positions on the device. Reflected signals were detected with a linear camera (S11639-01). For wide-field voltage mapping, an axicon (Thorlabs AX1205-B) and a rotating diffuser were used to generate incoherent Bessel illumination[21]. This allowed us to form a clean image, free of laser speckle. The Bessel ring was focused onto the back focal plane of the objective. Reflected-light images were acquired with a CMOS camera (Allied Vision Alvium 1800 U-052). Illumination was confined to a region of about $40 \times 40$ $\mu m^2$ around the channel, and the total power was kept below 60 nW, corresponding to an average intensity below 0.04 nW $\mu m^{-2}$ (3.8 mW $cm^{-2}$).

AC source and drain waveforms were generated by a dual-channel function generator (Siglent SDG6022X) with a fixed 180° phase offset. Each channel provided independent DC-offset and amplitude control to realize the calibration/biasing waveforms described in the main text. The source–drain current was inferred from the voltage drop across a 10 kΩ chip resistor in the source lead and measured with an SR830 lock-in amplifier synchronized by a TTL signal from the function generator. A TTL output triggered a DAQ to produce two camera triggers per period, capturing one image in the calibration half-cycle and one in the biasing half-cycle. Images from 10–40 periods were averaged to improve the signal-to-noise ratio. The drive frequency was 7.7 Hz.

**Bias-induced density spread**

The top gate and the sensor are held at fixed voltages during the measurement. A local target voltage $V$ therefore changes the local electron density of the target by $\delta n = -(C_{TG} + C_S)V/e = -\varepsilon_0\varepsilon_{hBN}(1/d_{TG} + 1/d_S)V/e$, where $C_{TG}$ and $C_S$ are the capacitances per unit area between the target and the top gate and between the target and the sensor, $\varepsilon_{hBN}$ is the out-of-plane dielectric constant of hBN, and $d_{TG}$ and $d_S$ are the thicknesses of the top-gate and interlayer hBN. For DV2 ($d_{TG}$ = 5.6 nm, $d_S$ = 3.8 nm) with $\varepsilon_{hBN} = 3$, $(C_{TG} + C_S)/e \approx 7.3 \times 10^{12}$ cm$^{-2}$ V$^{-1}$. Across the region used to extract $\Delta V$, the voltage varies by $\pm\Delta V/2$ around zero, and the density spans $n \pm (C_{TG} + C_S)\Delta V/2e$ (gray band in Fig. 4h). The corresponding conductivity change along the channel is estimated from the local slope of $\ln R_{\square}(n)$ at the same temperature.

## Data availability

The main data that support the findings of this study are available within the article and its Supplementary Information files. Source data are provided with this paper. More supporting data are available from the corresponding authors upon request.

## Acknowledgements

We gratefully acknowledge the contributions of all individuals involved in this research. We thank the members of our research group for helpful discussions.

## Funding Statement

This work is supported by the Director, Office of Science, Office of Basic Energy Sciences, Materials Sciences and Engineering Division of the US Department of Energy under contract number DE-AC02-05CH11231 (vdW heterostructure Program KCWF16). K.W. and T.T. acknowledge support from the CREST (JPMJCR24A5), JST and World Premier International Research Center Initiative (WPI), MEXT, Japan.

## Author contributions

F.W. and H.-L.K. proposed and designed the experiment; H.-L.K. and H.L. fabricated the devices with the help of R.X.; H.-L.K. and H.L. set up and performed optical measurements with input from R.Q. and C.U.; H.-L.K., H.L., and F.W. analyzed the data. Y.Y., H.-L.K., H.L., and F.W. performed local conductivity inversion. H.-L.K. fabricated and measured the reference transport device. K.W. and T.T. grew hBN crystals. H.-L.K. and F.W. wrote the manuscript with input from all authors.

## Competing interests

The authors declare no competing interests.

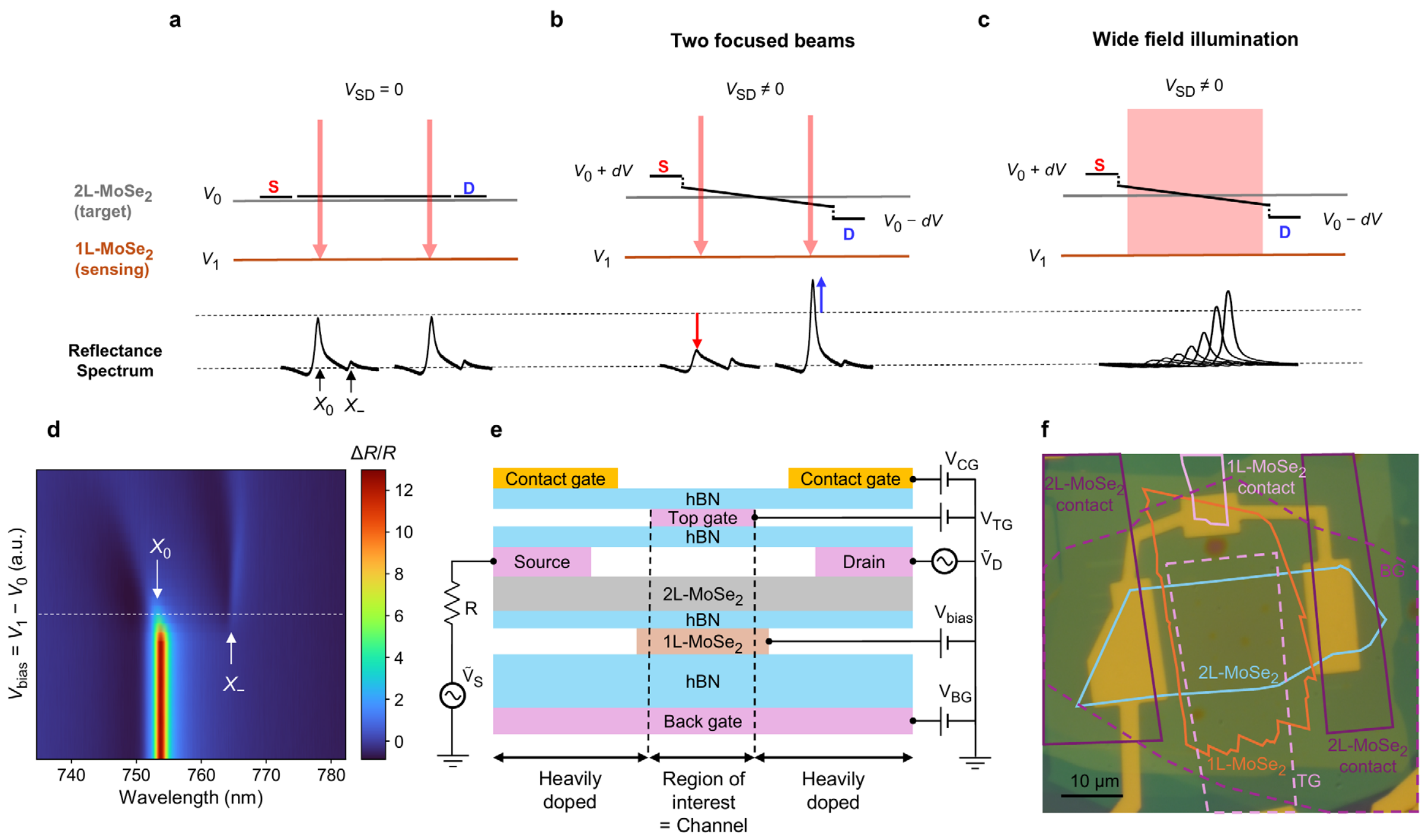

Fig. 1 | **Optical voltage profiling: principle and device architecture**. **a**, Zero-bias configuration (source–drain voltage $V_{\mathrm{SD}} = 0$) with the interlayer bias $V_{\mathrm{bias}} = V_1 - V_0$ set for maximum responsivity of the 1L-$MoSe_2$ sensor. $V_0$ and $V_1$ are the voltages of the target and the sensor, and 1L and 2L denote monolayer and bilayer. Focused beams (red arrows) probe discrete positions, analogous to voltage leads in a four-probe measurement. The 2L-$MoSe_2$ target electrostatically gates the sensor, modulating its carrier density and exciton resonance. **b**, Under finite source–drain bias ($V_{\mathrm{SD}} \neq 0$), current flows in the target. The voltage distribution shows large drops near the contacts and a linear gradient in the channel. Local changes in the sensor's exciton intensity (suppression or enhancement of the neutral exciton $X_0$) encode the target's local voltage. **c**, Wide-field illumination at 754 nm (tuned to the sensor exciton resonance) enables spatially resolved mapping of the target's local voltage across the region of interest. **d**, Reflection-contrast spectra ($\Delta R/R$) of the 1L-$MoSe_2$ sensor as a function of $V_{\mathrm{bias}}$, resolving the gate-dependent A-exciton $X_0$ and trion $X_-$ resonances. (White dashed line marks the bias value for optimal sensitivity). **e**, Cross-section of the heterostructure. Back-gate ($V_{BG}$) and contact-gate ($V_{CG}$) heavily dope the contact regions of 1L/2L-$MoSe_2$; the top-gate ($V_{TG}$) tunes the 2L-$MoSe_2$ channel while the interlayer bias ($V_{bias}$) is fixed for maximum responsivity. $\tilde{V}_{\mathrm{S}}$ and $\tilde{V}_{\mathrm{D}}$ are the AC voltages applied to the source and drain. **f**, Optical image of a representative device. Solid lines outline the 2L-$MoSe_2$ target (blue), the 1L-$MoSe_2$ sensor (orange) and the contacts to the target (purple) and the sensor (pink). Dashed lines outline the back gate (BG, purple) and the top gate (TG, pink).

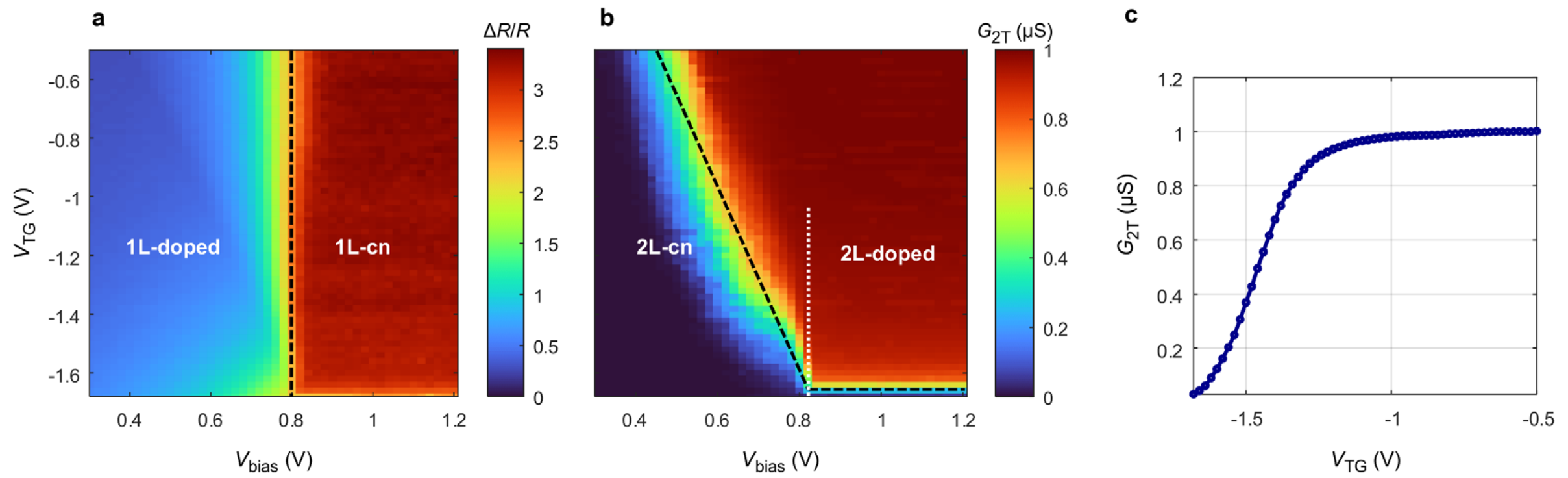

Fig. 2 | **Gate-dependent doping in the sensor and target layer**. **a,** Reflection contrast $\Delta R/R$ of the 1L-$MoSe_2$ (sensor) at the neutral-exciton ($X_0$) resonance as a function of top-gate voltage $V_{TG}$ and interlayer bias $V_{\mathrm{bias}}$. High $\Delta R/R$ marks the charge-neutral (cn) regime; low $\Delta R/R$ marks the electron-doped regime. The black dashed line marks the doping threshold of the sensor. Back-gate and contact gates are fixed at high positive voltages to render the contact regions metallic. **b,** Two-terminal conductance map $G_{2T}$ of the 2L-$MoSe_2$ (target) under a small AC bias (7.7 Hz, 10 mV). Regions of vanishing $G_{2T}$ indicate a charge-neutral regime, while a finite $G_{2T}$ indicates the electron-doped regime. The black dashed line marks the charge-neutrality line of the target, and the white dotted line marks the trajectory used in **c**. **c,** Line cut along the white dotted trajectory in **b**, taken at $V_{\mathrm{bias}} = 0.81$ V near the 1L sensor's phase boundary. Along this line, the sensor maintains its maximum sensitivity while the target electron density is varied across the metal–insulator transition. At large electron density, the two-terminal conductance saturates at around 1 μS due to a large contact resistance of ~1 MΩ.

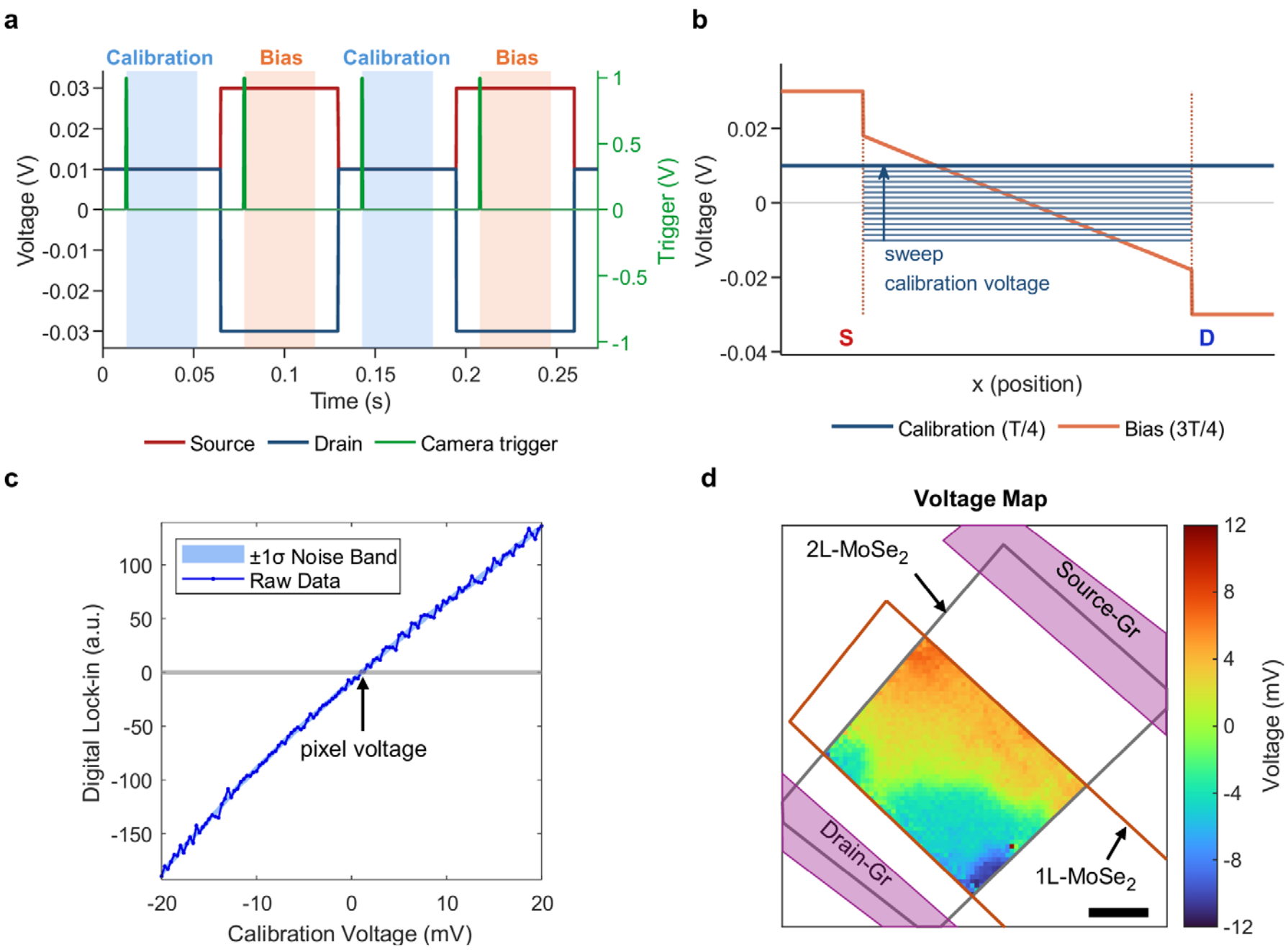

Fig. 3 | **In situ calibration and reconstruction of spatial voltage maps. a,** Time traces of the source (red) and drain (blue) voltages and the camera trigger (green) illustrating two phases per period: a calibration phase with $V_{\text{source}} = V_{\text{drain}} = V_{\text{cal}}$ (uniform voltage; no source–drain current) and a biasing phase with a small differential bias ($+30$ mV at the source and $-30$ mV at the drain). One image is acquired in each phase and frames are averaged over multiple periods. The calibration voltage $V_{\text{cal}}$ is stepped within ±20 mV while the biasing amplitude is held fixed. **b,** Position-dependent voltage $V(x)$ along the channel for several calibration voltages (blue) and for the biasing phase (orange), shown for the balanced-contacts case. For devices with contact imbalance, equal offsets are added to source and drain in the biasing phase to re-center the channel average near 0 V. **c,** Digital lock-in signal, $\Delta I(V_{\text{cal}}) = I_{\text{cal}}(V_{\text{cal}}) - I_{\text{bias}}$, where $I_{\text{cal}}(V_{\text{cal}}, x, y)$ is the image acquired during the calibration phase at voltage $V_{\text{cal}}$ and $I_{\text{bias}}(x, y)$ is the image acquired during the biasing phase. The signal is plotted versus the swept $V_{\text{cal}}$ at a representative pixel. The zero-crossing determines the local target-layer voltage. The shaded band indicates the ±1σ noise (3.1 a.u.), equivalent to 0.4 mV. **d,** Two-dimensional voltage map $V(x, y)$ in the 2L-$MoSe_2$ (DV1) with a bias current of 100 nA and a carrier density of $1.0 \times 10^{12}$ $cm^{-2}$. (Scale bar, 5 $\mu m$). The optically measured voltages are obtained in the area where the target (2L-$MoSe_2$) and sensor (1L-$MoSe_2$) layers overlap.

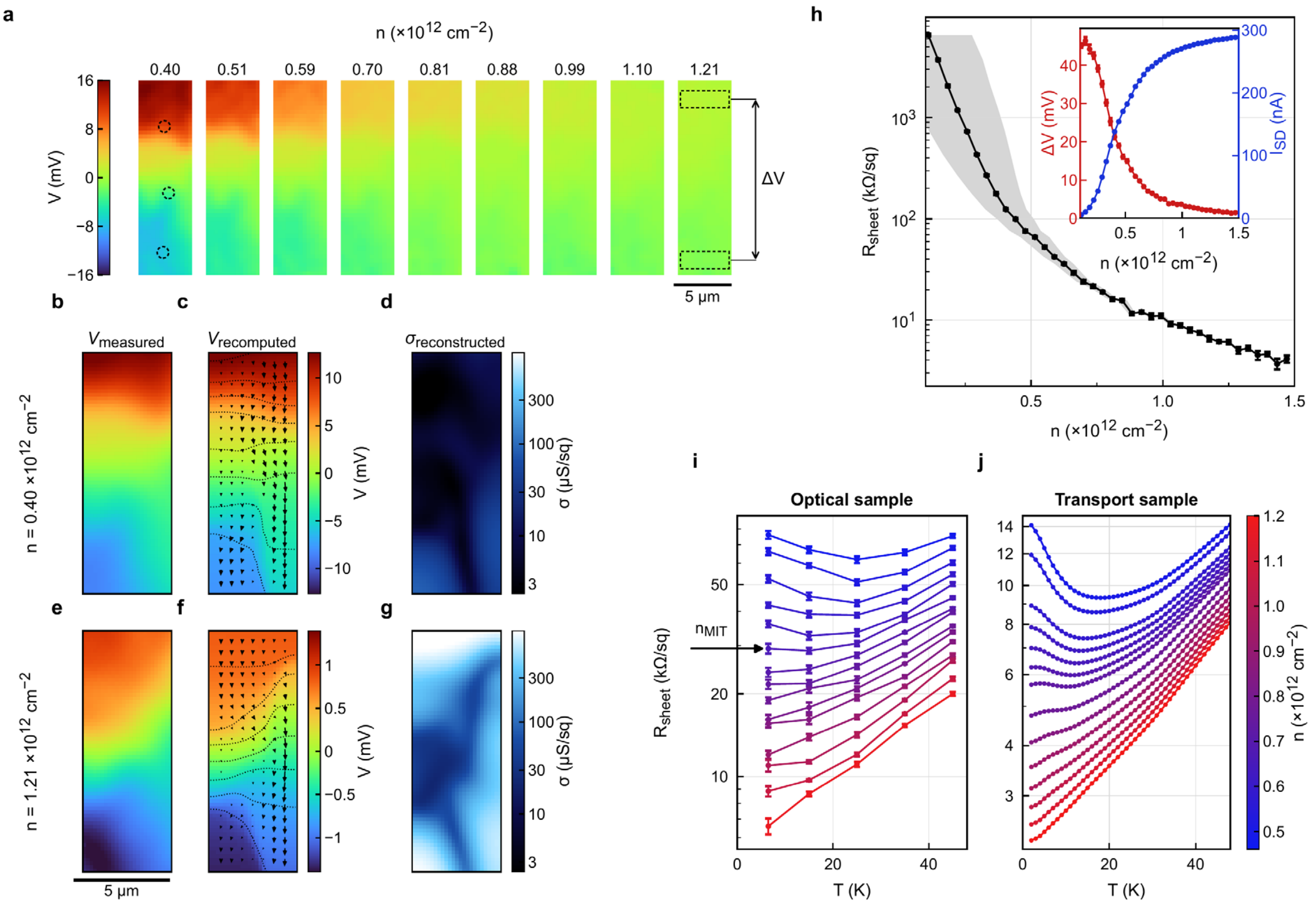

**Fig. 4 | Density-dependent voltage maps, local conductivity, and resistance across the metal–insulator transition (MIT) in 2L-MoSe₂. a**, Density-dependent spatial maps of the local voltage $V(x, y)$ for device DV2 measured at T = 6.5 K. Carrier densities (in units of $10^{12}$ cm$^{-2}$) are indicated above each panel. Dashed circles mark bubbles formed during heterostructure transfer. The source–drain bias voltage is 60 mV, and the corresponding bias current is shown in the inset of **h**. **b**, **e**, Measured voltage maps (Gaussian-smoothed) at carrier densities n = 0.40 × $10^{12}$ cm$^{-2}$ (**b**) and n = 1.21 × $10^{12}$ cm$^{-2}$ (**e**). **c**, **f**, Recomputed voltage maps obtained from the inverted local conductivity $\sigma$, together with the corresponding current density $\mathbf{J} = -\sigma\nabla V$. Arrows indicate the dominant current flow after thresholding weak contributions. **d**, **g**, Spatial maps of the reconstructed local conductivity at the same two densities. **h**, Average sheet resistance $R_\square$ versus carrier density. Inset: optically measured average four-probe voltage $\Delta V$ (red; difference between two interior regions marked on the n = 1.21 × $10^{12}$ cm$^{-2}$ map in **a**) and source–drain current $I_{\mathrm{SD}}$ (blue), from which $R = \Delta V / I_{\mathrm{SD}}$ is obtained. The gray band in **h** indicates the range of local carrier density within the analysis region, estimated from the source–drain voltage drop $\Delta V$ and the gate capacitances (Methods). **i**, $R_\square(T)$ from n = 0.5 to 1.2 × $10^{12}$ cm$^{-2}$ highlighting the crossover from insulating-like to metallic-like behavior with increasing density. **j**, $R_\square(T)$ measured on a separate 2L-MoSe₂ device by conventional four-probe transport over a similar density and temperature range. Error bars in **h** and **i** are estimated from the variation of $\Delta V$ among four parallel strips across the channel, each 1.1 µm wide to match the optical resolution, and from the fluctuation of $I_{\mathrm{SD}}$.

# Supplementary Information for

# Optical Voltage Profiling of 2D Semiconductors via Proximal Exciton Sensing

Ha-Leem Kim[1,2,3†,*], Hyungbin Lim[1,2,†], Yuanyi Yang[1], Ruishi Qi[1,2,3], Ruichen Xia[1], Can Uzundal[1,2,3], Takashi Taniguchi[4], Kenji Watanabe[5], Feng Wang[1,2,3,*]

[1]Department of Physics, University of California, Berkeley, CA 94720, USA.
[2]Materials Sciences Division, Lawrence Berkeley National Laboratory, Berkeley, CA 94720, USA.
[3]Kavli Energy NanoScience Institute, University of California, Berkeley and Lawrence Berkeley National Laboratory, Berkeley, CA 94720, USA.
[4]Research Center for Materials Nanoarchitectonics, National Institute for Materials Science, 1-1 Namiki, Tsukuba 305-0044, Japan.
[5]Research Center for Electronic and Optical Materials, National Institute for Materials Science, 1-1 Namiki, Tsukuba 305-0044, Japan.

## Supplementary Figures

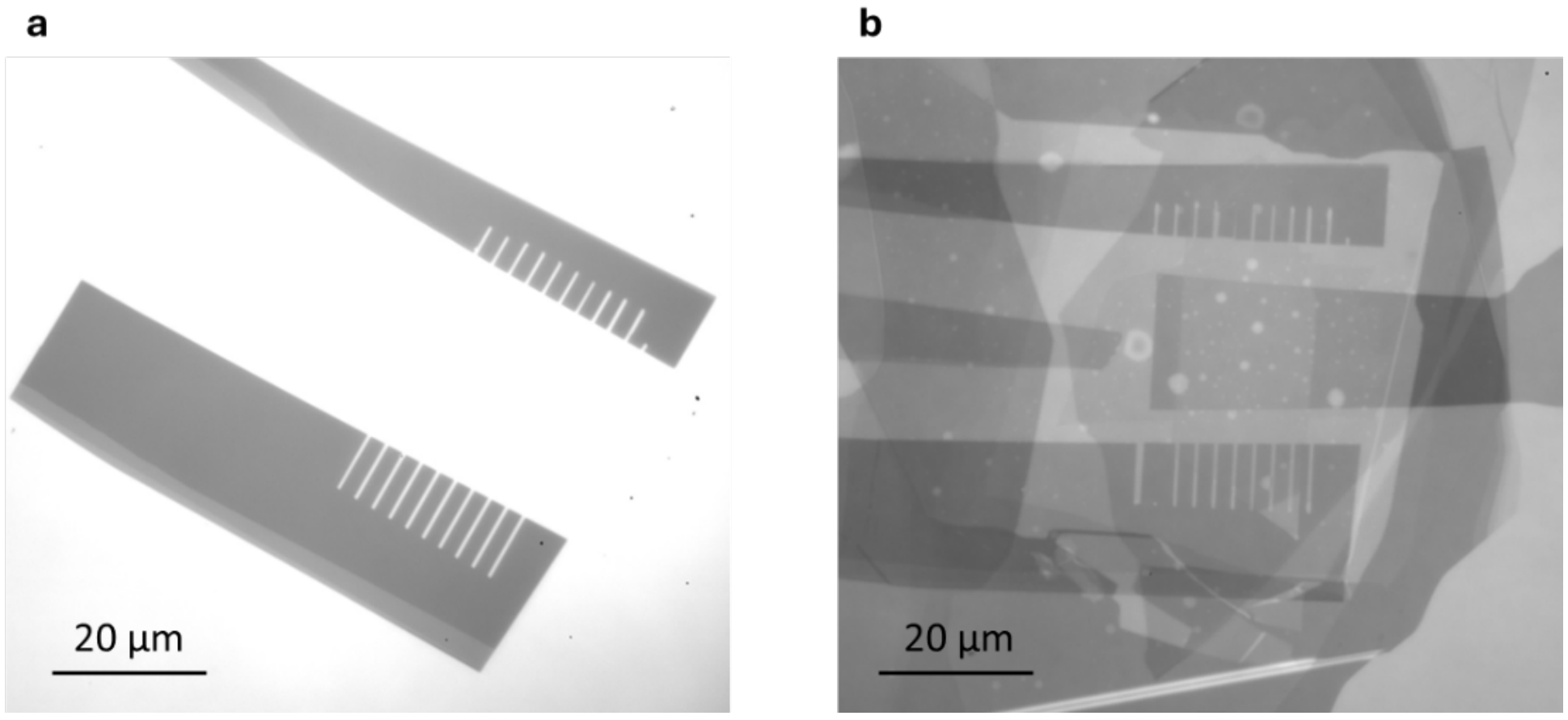

**Supplementary Figure 1 | Contact graphite for bilayer $MoSe_2$. a-b,** Optical microscope images of the graphite patterned into a comb geometry (a) and the assembled heterostructure (b).

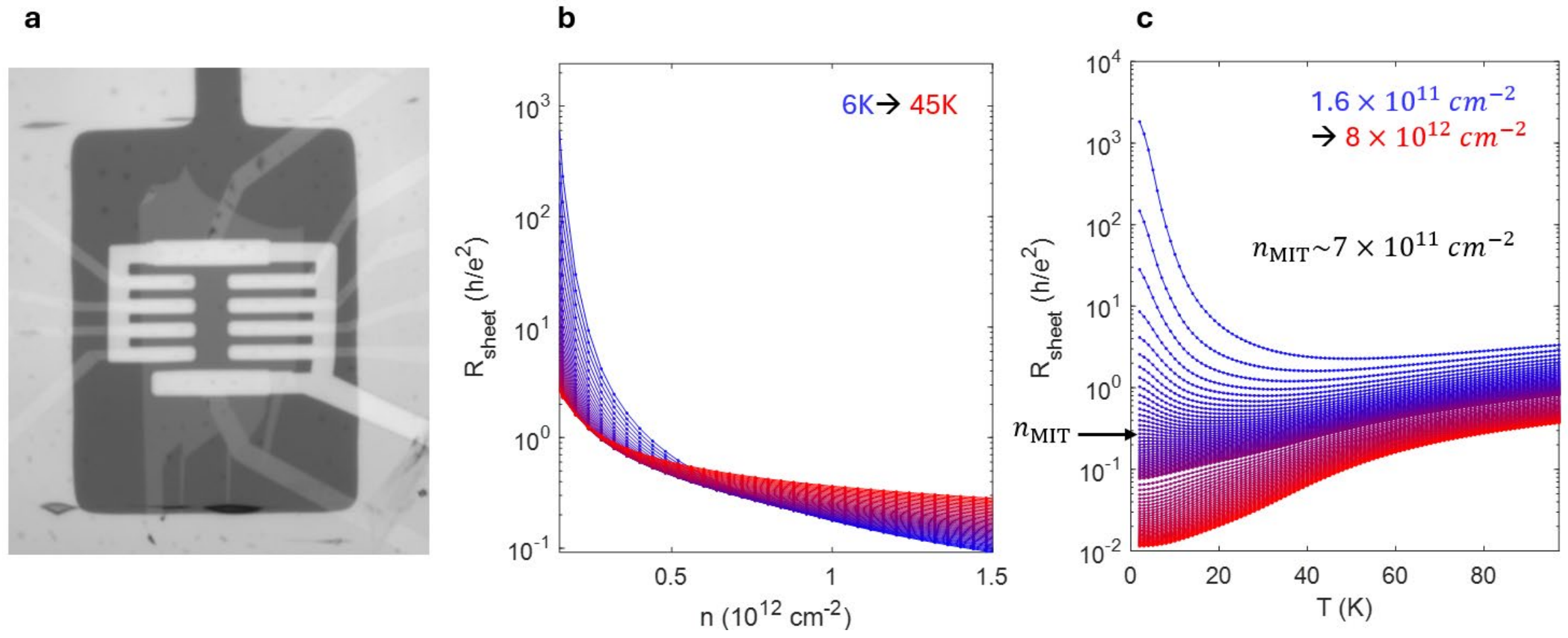

**Supplementary Figure 2 | Four-probe transport in bilayer MoSe₂. a**, Optical micrograph of the device. **b**, Sheet resistance $R_\square$ versus carrier density at temperatures from 6 to 45 K. **c**, Temperature dependence of $R_\square(T)$ at carrier densities from $1.6 \times 10^{11}$ to $8 \times 10^{12}$ cm$^{-2}$. The arrow marks $R_\square \approx 0.23\ h/e^2$ (≈6 kΩ) at the critical density $n_{\mathrm{MIT}} \approx 7 \times 10^{11}$ cm$^{-2}$, where $\mathrm{d}R_\square/\mathrm{d}T$ changes sign at low temperature.

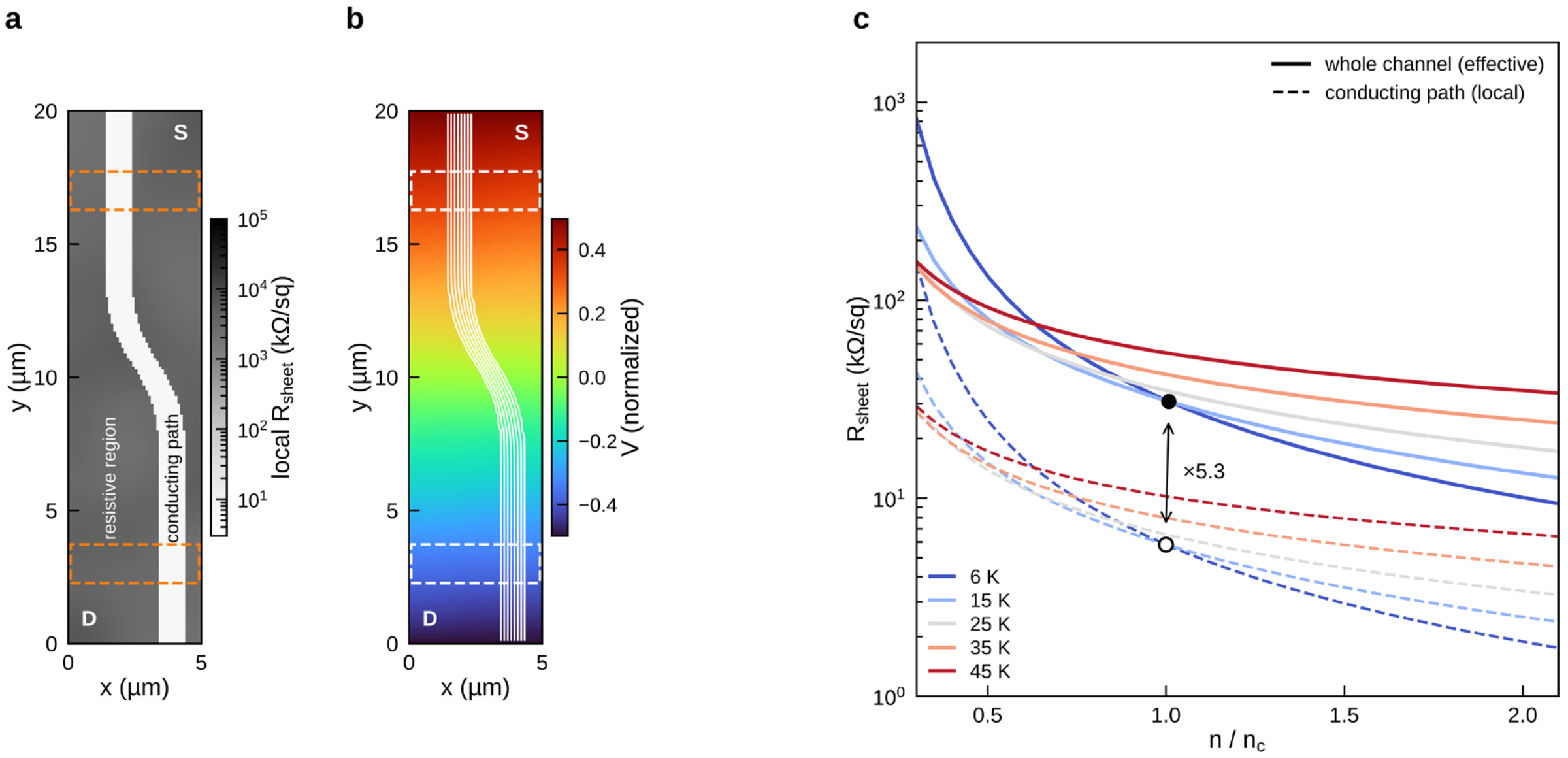

**Supplementary Figure 3 | Simulated effect of a narrow conducting path on the extracted sheet resistance. a**, Local sheet resistance at $n = n_c$ and $T$ = 6 K, where $n_c$ is the critical density of the conducting path. The path, one-fifth of the channel width, runs from the source (S) at the top to the drain (D) at the bottom through a resistive region. Orange dashed lines mark the two voltage-averaging bands, 1.44 µm thick and 14 µm apart. **b**, Voltage map and current flow lines (white) for the same conductivity map. Adjacent lines enclose equal current. **c**, $R_\square$ versus $n/n_c$ from 6 to 45 K. The path follows the $R_\square(n,T)$ of the transport device (Supplementary Fig. 2b), with a critical resistance of 5.8 kΩ. The resistive region follows the same curves shifted to a 2.2 times higher density, with random variations of about 7% on a 1-µm scale, and scaled up 100 times, giving about 3 MΩ/sq at $n_c$ and 6 K. Solid lines: effective sheet resistance of the whole channel, $R_\square = (\Delta V/14\ \mu\mathrm{m})/(I/W)$, with $W$ the channel width. Dashed lines: local sheet resistance of the conducting path. Circles mark the crossing of the 6 K and 15 K curves.

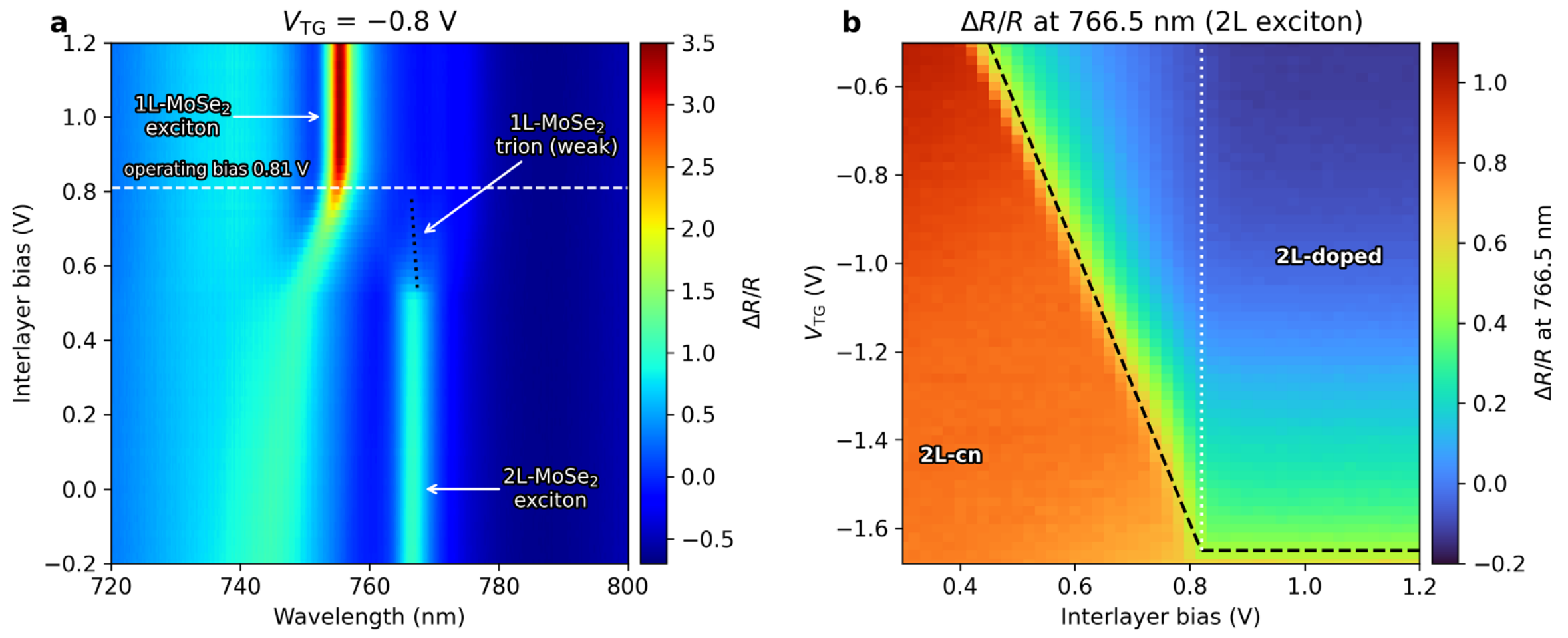

**Supplementary Figure 4 | Reflection-contrast spectra of the sensor and target.** Device DV1 at 6.5 K. **a**, $\Delta R/R$ as a function of wavelength and interlayer bias at $V_{\mathrm{TG}}$ = −0.8 V. Arrows mark the excitons of the 1L-$MoSe_2$ sensor (754 nm) and the 2L-$MoSe_2$ target (766.5 nm); the black dotted line follows the weak sensor trion; the horizontal white dashed line marks the operating bias of 0.81 V. **b**, $\Delta R/R$ at 766.5 nm as a function of interlayer bias and $V_{\mathrm{TG}}$. The black dashed line is the charge-neutrality boundary of the target from the two-terminal conductance (Fig. 2b of the main text); the white dotted line is the voltage-profiling trajectory at $V_{\mathrm{bias}}$ = 0.81 V.

# Supplementary Notes

## Supplementary Note 1. AFM nanolithography of graphite contact

Graphite contacts to bilayer $MoSe_2$ often yield uncontrolled contact resistances. Some devices exhibit resistances above 100 MΩ, while others reach values as low as ~ 100 kΩ. Since charge injection occurs along the edge of the graphite/2L-$MoSe_2$ interface, extending the contact boundary increases the probability of forming low-resistance interfaces. We patterned the graphite into a comb shape using AFM nanolithography[1] (Supplementary Fig. 1a), thereby extending the effective edge length at the 2L-$MoSe_2$/graphite interface. This comb geometry reproducibly gives contact resistances as low as ~400 kΩ. Supplementary Fig. 1b shows a representative device image incorporating the comb-shaped graphite contact.

## Supplementary Note 2. Voltage resolution

The local voltage at each pixel is obtained from the zero-crossing of the digital lock-in signal $\Delta I(V_{\mathrm{cal}})$ (Fig. 3c of the main text). Near the zero-crossing, $\Delta I$ varies linearly with $V_{\mathrm{cal}}$, and a fluctuation of the signal shifts the zero-crossing by that fluctuation divided by the slope $|\partial \Delta I / \partial V_{\mathrm{cal}}|$. A linear fit over $N$ calibration steps centered on the zero-crossing averages the independent noise of these steps, giving

$$\sigma_V = \frac{\sigma_{\Delta I}}{|\partial \Delta I / \partial V_{\mathrm{cal}}|\sqrt{N}},$$

where $\sigma_{\Delta I}$ is the standard deviation of $\Delta I$ for a single calibration step. With the fit window centered on the zero-crossing, the uncertainty of the fitted slope does not contribute to first order. For the pixel in Fig. 3c, $\sigma_{\Delta I} = 3.1$ a.u. and $\partial \Delta I / \partial V_{\mathrm{cal}} = 7.7$ a.u. mV$^{-1}$, corresponding to a voltage noise of 0.4 mV for a single calibration step. A fit window of ±3 mV contains $N = 19$ steps and yields $\sigma_V \approx 0.09$ mV for 20 periods of averaging per calibration step (about 2.6 s at 7.7 Hz).

## Supplementary Note 3. Local conductivity inversion via finite-element forward–adjoint optimization

We reconstruct the local sheet conductivity σ(x, y) within a rectangular region of interest (ROI) from the measured potential map $V_{meas}(x, y)$. Current is taken to flow along the y-axis. The ROI is the open rectangle Ω = (0, W) × (0, L), where W is the ROI width (perpendicular to current flow) and L is the ROI length along the source–drain (current-flow) direction; both are taken in physical (μm) units. The boundary ∂Ω is partitioned into the inflow edge $\Gamma_{in}$ (source contact at y = L, where current enters), the outflow edge $\Gamma_{out}$ (drain contact at y = 0, where current exits), and the two long edges at x = 0 and x = W (no-flux walls parallel to the y-axis). The reconstruction is formulated as a PDE-constrained optimization governed by steady-state charge conservation and Ohm's law, and is solved by a forward–adjoint method. Because the inverse problem is ill-posed — many σ fields can produce nearly identical $V_{meas}$ — we add a smoothness penalty that suppresses unphysical (jagged or oscillatory) σ fields and let the optimization converge toward the most physically reasonable solution within the data-consistent family. Rather than entering as a strict penalty term in a fixed objective, this regularization is applied as a bias on the descent direction at each step (see Algorithm). All finite-element (FE) operators are assembled directly in MATLAB.

To improve numerical conditioning, all internal computations are nondimensionalized using a reference length $L_{ref} = \sqrt{(W\,L)}$ (the geometric mean of the ROI's two physical dimensions), a reference potential $V_{ref} = \max|V_{meas}|$, and a reference conductivity $\sigma_{ref} = I_{rect} / V_{ref}$, where $I_{rect}$ is the total current carried by the ROI (computed from the measured device current scaled by the ROI/device width ratio, assuming a uniform current density across the device width). All physical quantities are mapped to dimensionless form before assembly, and rescaled back to physical units at output.

### Forward problem

The ROI is discretized as a structured rectangular triangulation, with each pixel cell of $V_{meas}$ split into two right triangles. Both the unknown conductivity σ and the potential u(x, y) are represented as first-order (P1) Lagrange fields — piecewise-linear functions defined on the triangulation, with one nodal basis function per mesh vertex (each basis function takes the value 1 at its own vertex and 0 at every other vertex, and is linear within each triangle). Within an element e, the elemental conductivity $\sigma_e$ is the centroid average of its three vertex values. Assuming isotropic local conductivity and stationary transport, u satisfies

$$-\nabla \cdot (\sigma\, \nabla u) \;=\; 0 \quad in\, \Omega,$$

with current-injection (Neumann) boundary conditions on the source/drain edges and zero-flux on the long edges:

$$\sigma\, \nabla u \;\cdot\; n \;=\; \pm\, \frac{I_{rect}}{W_{rect}} \quad on\;\Gamma_{in} \,/\, \Gamma_{out}\,, \qquad \sigma\, \nabla u \;\cdot\; n \;=\; 0 \quad on\;\partial\Omega \setminus (\Gamma_{in} \,\cup\, \Gamma_{out}).$$

Here n is the outward unit normal on ∂Ω; $W_{rect}$ is the ROI width along the contact edge (numerically equal to W); and the sign on $\Gamma_{in}$ / $\Gamma_{out}$ is fixed by comparing row-averaged $V_{meas}$ between the top (y = L) and bottom (y = 0) edges (auto-detected so the algorithm is agnostic to current polarity). Because the pure-Neumann problem is defined only up to an additive constant, the gauge is fixed by pinning a single Dirichlet reference DOF — the node with the smallest $|V_{meas}|$ — to u = 0. For multi-density sweeps, optional soft anchors (a regular interior grid of nodes whose data-misfit weight is increased by a factor $w_{anchor} \approx 5$) stabilize the gauge across runs without forcing local σ artifacts.

### Inverse problem

Two sources of ill-posedness are addressed separately. (i) An overall amplitude ambiguity from the uncertain effective current density at the boundary is removed by a single scalar amplitude factor $C^*$ fit by least squares between the computed potential $u_c = u(\sigma)$ and $V_{meas}$:

$$C^* \;=\; argmin_C\, (\|C\, u_c \;-\; V_{meas}\|)^2 \;=\; \frac{\langle u_c, V_{meas}\rangle}{\langle u_c, u_c\rangle},$$

where ‖·‖ denotes the $L^2(\Omega)$ norm and ⟨·,·⟩ the corresponding inner product (discretized as nodal dot products with the FE mass matrix). The conductivity is then rescaled as $\sigma \leftarrow \sigma / C^*$ — division, not multiplication. The direction follows from the linearity of the forward problem in $\sigma^{-1}$: for fixed boundary current, u(σ) scales as $\sigma^{-1}$, so $u(\sigma / C^*) = C^*\, u(\sigma) = C^*\, u_c \approx V_{meas}$, i.e., dividing σ by $C^*$ makes the model match the measured amplitude exactly. (ii)

Spatial under-determination is addressed via regularization (described below). The data-misfit functional is the weighted $L^2$ residual,

$$J_{data}(\sigma) = \frac{1}{2}\int_{\Omega} w(x)\,(C^*\,u(\sigma) - V_{meas})^2\,d\Omega,$$

with the misfit weight $w(x) = 1$ by default, elevated to $w_{anchor}$ at soft-anchor nodes. Two regularization functionals are evaluated,

$$R_{Tik}(\sigma) = \frac{1}{2}\alpha_{Tik}\int_{\Omega} |\nabla\sigma|^2\,d\Omega\,, \quad R_{TV}(\sigma) = \alpha_{TV}\int_{\Omega}\sqrt{|\nabla\sigma|^2 + \varepsilon^2}\,d\Omega\,,$$

with Tikhonov weight $\alpha_{Tik}$, total-variation weight $\alpha_{TV}$, and TV smoothing parameter $\varepsilon$ (set to a small constant to keep the integrand differentiable near $|\nabla\sigma| = 0$). These functionals shape the descent direction rather than being strictly minimized as a penalty (see Algorithm).

### Forward–adjoint gradient and adaptive (κ-scaled) descent

At each iteration, given the current $\sigma$ and forward solution u, the adjoint potential p is the solution of $K(\sigma)\,p = -M_w\,r$ with the same Dirichlet pin as the forward problem. Here $K(\sigma)$ is the conductivity-weighted FE stiffness matrix, $M_w$ is the diagonal mass matrix carrying the misfit weights w, and $r = C^*\,u - V_{meas}$ is the nodal residual. The data-misfit gradient at vertex i is then assembled by Galerkin projection over all triangles e sharing that vertex,

$$\left(\frac{\partial J_{data}}{\partial\sigma}\right)_i = \Sigma_{e\,\ni\,i}\,\frac{A_e}{3}\,(\nabla u_e\cdot\nabla p_e),$$

where $A_e$ is the area of triangle e, and $\nabla u_e$, $\nabla p_e$ are the (constant) gradients of the P1 fields on that element. The regularization gradient is the $L^2$-projected (mass-inverted) weak form of the variational derivative of $R_{Tik} + R_{TV}$. Because the data and regularization gradients have different physical scales, an adaptive rescaling is applied at each iteration so that $\|\partial R/\partial\sigma\|_{scaled} = \kappa\,\|\partial J_{data}/\partial\sigma\|$, with $\kappa = 0.1$. The values of $\alpha_{Tik}$ and $\alpha_{TV}$ thus control only the Tikhonov-vs-TV character of the regularization gradient direction; κ is the effective smoothness-strength knob. A cubic continuation, $\alpha_{eff} = \alpha\cdot(\min(1, k/T_{ramp}))^3$ with iteration index k and ramp length $T_{ramp} = 50$, ramps the regularization in over the first $\approx T_{ramp}/2$ iterations to keep the early steps data-driven.

Optimization is performed in $m = \ln\sigma$ to keep $\sigma$ positive, with descent direction

$$d_m = -\frac{\sigma\odot\nabla J_{total}}{\|\sigma\odot\nabla J_{total}\|}\,,\quad \nabla J_{total} = \partial J_{data}/\partial\sigma + (\partial R/\partial\sigma)_{scaled}.$$

Here $\odot$ denotes the Hadamard (elementwise) product, and the chain rule factor $\sigma\,\odot$ converts the gradient with respect to $\sigma$ into one with respect to m. Step length is chosen by Armijo backtracking, with the sufficient-decrease criterion applied to the data-misfit functional only (not to the combined $J = J_{data} + R$):

$$J_{data}(m + \alpha\,d_m) \le J_{data}(m) + c_1\,\alpha\,\langle d_m, \partial J_{data}/\partial m\rangle\,,\quad c_1 = 10^{-5}\,, \alpha_0 = 2\,,\quad \rho = 0.5\,.$$

with sufficient-decrease constant $c_1$, initial trial step $\alpha_0$, and backtracking factor $\rho$. This data-only acceptance rule allows $R_{Tik}$ and $R_{TV}$ to vary freely while the κ-scaling caps their relative gradient norm. The role of regularization in this scheme is therefore to bias the descent direction toward smoother $\sigma$ at each step, not to impose a fixed minimization of a regularized objective. The recovered $\sigma$ is robust to changes in $\alpha_{Tik}$ and $\alpha_{TV}$ at fixed κ, since these affect only the regularization gradient direction.

## Supplementary Note 4. Conventional four-probe transport measurement on bilayer $MoSe_2$

A dedicated transport device was fabricated to probe the metal–insulator transition in bilayer (2L) $MoSe_2$. The back gate was patterned by photolithography and metal evaporation (Cr/Pt, 2 nm/8 nm). After lift-off, the patterned back-gate surface was cleaned by AFM contact-mode scanning, and a 50 nm hBN flake was transferred. The hBN surface was then AFM-cleaned prior to depositing Pt contact electrodes (8 nm Pt). A 5 nm upper hBN and the 2L-$MoSe_2$ flake were picked up using a PC (polycarbonate) film and released onto the pre-patterned Pt contacts. Interlayer bubbles formed during the transfer process were squeezed away by slow AFM contact-mode scans (~12 h total). Finally, a contact gate was defined by e-beam lithography and metal deposition (Cr/Au, 5 nm/60 nm). An optical image of the completed device is shown in Supplementary Fig. 2a. Resistance was measured in a variable-temperature insert (VTI) cryostat (C-Mag Vari-12T, base $T = 1.8$ K). Supplementary Fig. 2b displays sheet resistance $R_{\square}(n)$ at multiple temperature points spanning 6 to 45 K, revealing a metal–insulator crossover. We identify a critical density $n_{\mathrm{MIT}} \approx 7 \times 10^{11}$ cm$^{-2}$ from the sign change of $\mathrm{d}R_{\square}/\mathrm{d}T$ at low temperature (Supplementary Fig. 2c). At $n_{\mathrm{MIT}}$, $R_{\square} \approx 6$ kΩ (0.23 $h/e^2$). Supplementary Fig. 2c displays $R_{\square}(T)$ at multiple density points spanning $1.6 \times 10^{11}\ cm^{-2}$ to $8.0 \times 10^{12}\ cm^{-2}$. The extracted $n_{\mathrm{MIT}}$ is consistent with the value obtained from the optical voltage-profiling measurements in the main text.

The critical resistance of the transport device is about five times lower than that of DV2, whereas the critical densities agree. The heterostructure of DV2 contains more than ten transferred layers, for which removing bubbles by AFM scanning is not practical. Supplementary Fig. 3 illustrates how a narrow conducting path affects the extracted sheet resistance. In this simple phenomenological simulation, a conducting path one-fifth of the channel width runs from the source to the drain and has the $R_{\square}(n, T)$ of the transport device. The rest of the channel represents regions affected by bubbles and strain and is much more resistive than the path near $n_{\mathrm{MIT}}$. We solve $\nabla \cdot (\sigma \nabla V) = 0$ and obtain $R_{\square}$ from the voltage difference between two bands 14 µm apart and the total current, as in Fig. 4h of the main text. For the whole channel, the crossover occurs within 1% of the critical density of the conducting path, but the critical resistance is 5.3 times that of the path, close to the ratio of the channel width to the path width. A narrow conducting path thus leaves $n_{\mathrm{MIT}}$ unchanged while increasing the critical resistance. In a real device, the situation is more complex. The conductivity contrast between the conducting path and the rest of the channel depends on density and can decrease at higher doping. Consistent with a decreasing contrast, the ratio of the sheet resistances of the two devices decreases from about 4–5 near $n_{\mathrm{MIT}}$ to about 2 at $1.5 \times 10^{12}$ cm$^{-2}$.

### Supplementary Note 5. Spectral separation of the sensor and target resonances

Supplementary Fig. 4a shows reflection-contrast spectra of device DV1 as a function of interlayer bias at $V_{TG} = -0.8$ V. The neutral excitons of the 1L-$MoSe_2$ sensor and the 2L-$MoSe_2$ target lie at 754 nm and 766 nm, and a weak feature at 765–770 nm between the doping onset of the target and the sensor threshold is the trion of the sensor. The feature at 766 nm is the exciton of the target rather than the sensor trion. Its intensity over the full gate space (Supplementary Fig. 4b) is bounded by the charge-neutrality line of the target obtained from the two-terminal conductance (Fig. 2b of the main text), not by the sensor threshold at $V_{bias} \approx 0.8$ V. The diode laser used for voltage profiling is temperature-stabilized on a single longitudinal mode near 754 nm, with no mode hopping during the measurements, and its emission width is far below 1 nm. The target exciton lies more than 10 nm from the probing wavelength. Any residual contribution of the target to the reflectance at 754 nm is included in the pixel-wise in situ calibration, since every measurement period contains a calibration phase and a biasing phase under the same gate setting (Fig. 3a of the main text).

## Supplementary References